\documentclass[twocolumn]{aastex631}

\usepackage{amsmath}

\newcommand{\hi}{\hbox{H\,{\sc i}}}
\newcommand{\mhi}{\hbox{${M}_{\rm HI}$}}
\newcommand{\mhiin}{\hbox{${M}_{\rm HI,in}$}}
\newcommand{\mhiout}{\hbox{${M}_{\rm HI,out}$}}
\newcommand{\rhi}{R_{{\rm HI}}}

\newcommand{\dmhi}{\hbox{$\Delta {M}_{\rm HI}$}}
\newcommand{\dmhiin}{\hbox{$\Delta {M}_{\rm HI,in}$}}
\newcommand{\dsfr}{\hbox{$\Delta {\rm SFR}$}}
\newcommand{\dz}{\hbox{$\Delta Z$}}

\newcommand{\ha}{\hbox{H$\alpha$}}
\newcommand{\hb}{\hbox{H$\beta$}}

\newcommand{\nii}{\hbox{[N\,{\sc ii}]}}

\newcommand{\oiii}{\hbox{[O\,{\sc iii}]}}

\newcommand{\Msun}{\hbox{$M_{\sun}$}}

\shorttitle{MZR with $\hi$}
\shortauthors{Chen et al.}

\graphicspath{{./fig/}}

\begin{document}

\title{The Role of Inner $\hi$ Mass in Regulating the Scatter of the Mass--Metallicity Relation}
\correspondingauthor{Jing Wang}
\email{jwang\_astro@pku.edu.cn}

\author[0000-0002-5016-6901]{Xinkai Chen}
\affiliation{CAS Key Laboratory for Research in Galaxies and Cosmology, Department of Astronomy, University of Science and Technology of China, Hefei 230026, People’s Republic of China}
\affiliation{School of Astronomy and Space Science, University of Science and Technology of China, Hefei 230026, People’s Republic of China}

\author[0000-0002-6593-8820]{Jing Wang}
\affiliation{Kavli Institute for Astronomy and Astrophysics, Peking University, Beijing 100871, China}

\author[0000-0002-7660-2273]{Xu Kong}
\affiliation{CAS Key Laboratory for Research in Galaxies and Cosmology, Department of Astronomy, University of Science and Technology of China, Hefei 230026, People’s Republic of China}
\affiliation{School of Astronomy and Space Science, University of Science and Technology of China, Hefei 230026, People’s Republic of China}
\affiliation{Frontiers Science Center for Planetary Exploration and Emerging Technologies, University of Science and Technology of China, Hefei, Anhui, 230026, China}

\begin{abstract}
We use 789  disk-like, star-forming galaxies (with 596 $\hi$ detections) from  $\hi$ follow-up observations for the SDSS-IV MaNGA survey to study the possible role of inner $\hi$ gas in causing secondary dependences in the mass--gas-phase metallicity relation. We use the gas-phase metallicity derived at the effective radii of the galaxies.
We derive the inner $\hi$ mass within the optical radius, but also use the total $\hi$ mass and star formation rate (SFR) for a comparison. We confirm the anticorrelation between the total $\hi$ mass and gas-phase metallicity at fixed stellar mass, but the anticorrelation is significantly strengthened when the total $\hi$ mass is replaced by the inner $\hi$ mass. 
Introducing a secondary relation with the inner $\hi$ mass can produce a small but noticeable decrease (16\%) in the scatter of the mass--gas-phase metallicity relation, in contrast to the negligible effect with the SFR. The correlation with the inner $\hi$ mass is robust when using different diagnostics metallicity, but the correlation with SFR is not. The correlation with the inner $\hi$ mass becomes much weaker when the gas-phase metallicity is derived in the central region instead of at the effective radius. 
These results support the idea that the scatter in the mass--metallicity relation is regulated by gas accretion, and not directly by the SFR, and stress the importance of deriving the gas mass and the metallicity from roughly the same region. The new relation between inner $\hi$ mass and gas-phase metallicity will provides new constraints for chemical and galaxy evolution models.

\end{abstract}



\section{Introduction}
Neutral atomic hydrogen ($\hi$) gas is the raw material from
which molecular gas and then stars form. Revealing and quantifying its relationship to other galactic properties thus provides key constraints for galaxy evolution models, particularly the processes directly involving gas, including gas accretion, star formation (SF), and feedback \citep{somerville2015physicala,naab2017theoreticala}. Over the past decade, single-dish radio telescopes have provided  statistically large samples of $\hi$ gas, giving us a chance to explore those relationships. For example, the $\hi$ mass fraction has been found to be correlated  with the color, specific star formation rate (SFR), and effective  stellar-mass surface density of galaxies \citep{catinella2010galex,catinella2018xgass,chen2020growth}.
However, as $\hi$ is typically much more radially extended than the stellar disk in star-forming galaxies \citep[SFGs;][]{swaters2002westerbork, wang2014observational}, the lack of spatial information in single-disk data limits the physical interpretation of the observed $\hi$ relations. To mitigate this problem, \cite{wang2020xgass} have recently introduced a method to separate the total $\hi$ masses into inner and outer parts, with a radial division near the edge of the optical disk. The method makes use of the tight relation between the $\hi$ mass and the characteristic $\hi$ radius $R_{\rm HI}$ \citep{broeils1997short, swaters2002westerbork, noordermeer2005westerbork, wang2014observational, wang2016new}, and the universal shape of the $R_{\rm HI}$-normalized $\hi$ radial distribution in the outer disks \citep{swaters2002westerbork,wang2014observational, wang2016new}. 
With the newly derived inner $\hi$ mass, we can have a new perspective on galaxy evolution. For example, \cite{wang2020xgass} showed that the spread along the star forming main sequence (SFMS) is best characterized by the averaged inner $\hi$ mass surface density among $\hi$-related parameters. They also showed that the offset of disk galaxies from the SFMS correlates with the conversion efficiency of the inner $\hi$ to molecular gas; such a trend is not seen when the total, rather than the inner $\hi$, masses are considered. It will be interesting and important to link the inner $\hi$ masses to other key parameters that provide clues to galaxy evolution. 

Gas-phase metal abundance is one of the most important  components for understanding the evolution of galaxies. It was found from long ago that there is a tight relation between the gas-phase metallicity ($Z$) and the stellar mass of galaxies \citep{lequeux1979chemical, tremonti2004origin}, which has a scatter of only $\sim$0.1 dex. Such a relation is commonly referred to as the MZR. A longer evolutionary history, a deeper gravitational potential, which prevents metal loss due to outflows, and a lower richness in cold gas have been considered as the three major reasons for high-mass galaxies to have a higher $Z$ than low-mass galaxies \citep[e.g.][]{tremonti2004origin,maiolino2008amaze,chisholm2018metalenriched}.

Most of the metals observed at present may be produced in early periods of galaxy evolution. Explorations of the evolutions of metals in galaxies suggest that the enrichment in galaxies more massive than $10^{9.5}~M_{\odot}$ occurs very early in the past of those galaxies \citep[e.g.][]{valeasari2009evolution, camps-farina2021evolution}. This is because after a single episode of SF, the production of oxygen reaches its saturated value very fast \citep{maiolino2019re}. However, the accretion of metal poor gas from the circumgalactic medium or low-mass satellite galaxies can play an important role in reshaping the existing metallicity, by diluting it. Such an effect has been extensively studied by means of gas regulator models, which consist of a series of continuity equations under the assumption of instantaneous recycling \citep{dave2012analytic, lilly2013gas}. 
In the context of $\Lambda$ cold dark matter cosmology (which sets the gas accretion model; \citealp{mo1998formation}), with constraints from observations (e.g., the SF law; \citealp{schmidt1959rate}), most gas regulator models have reached the conclusion that not only does gas accretion play a key role in diluting the gas-phase metallicity, but the observed MZR is also useful in constraining gas accretion models \citep[e.g.][]{wang2021gasphase}. However, gas regulator models that are radially resolved, and consider the radial inflow of cool gas, have only been investigated recently \citep[e.g.][]{sharda2021physics}. A close comparison of these relatively new models with the observed radial properties of $\hi$ is still largely missing. 
After gas accretion, SF can be elevated as a byproduct of enhanced gas richness \citep{wang2020xgass} and eventually lead to metal enrichment, but the dilution due to the newly accreted gas may last for a long time before the gas is significantly depleted, because a significant fraction of $\hi$ is stored in the extended outer disk \citep{swaters2002westerbork, wang2014observational}, far beyond the reach of immediate SF, and gradually flows radially inward \citep{krumholz2018unified}.

So far, as predicted by the gas regulator models, observations have revealed the secondary dependence of the MZR on molecular gas mass \citep{bothwell2016galaxy} and $\hi$ mass \citep{bothwell2013fundamental, brown2018role,zu2020gas}. As the gas diluting the metal also fuels the SF, a secondary dependence of the MZR on the SFR has also been found in observations \citep[e.g.][]{ellison2008clues,mannucci2010fundamental}, which, however, seems to be weak and only observable at low masses in recent studies \citep{sanchez2019sami}. The stronger dependence of the MZR on neutral gas mass than on SFR suggests that the latter is only an indirect manifestation of the former, when linking to the process of accreted low-$Z$ gas diluting the metallicity. Although there seems to be observational evidence for $\hi$ regulating the scatter of the MZR \citep{brown2018role}, the relevant studies had to rely on low-resolution $\hi$ data, which had the problem of inconsistent radial extension between the investigated properties, as we described earlier. In extreme cases, the metallicity is derived from the central region (e.g. the 3$''$ fiber region of the Sloan Digital Sky Survey, or SDSS; \citealp{york2000sloan}), while the $\hi$ extends four times farther beyond the optical disk for an SFG  \citep{swaters2002westerbork, wang2014observational}.

The Mapping Nearby Galaxies at Apache Point Observatory (MaNGA) survey \citep{bundy2015overview}, and its newly released $\hi$ follow-up (HI-MaNGA;\citealt{masters2019imanga,stark2021imanga}),  allow us to revisit the relation between $\hi$ masses and gas-phase metallicity, by measuring these quantities from roughly the same region. 
The $\hi$ data still constitute integral spectra of galaxies, but new developments in analysis techniques make it possible to estimate the $\hi$ mass within the optical radius with relatively small scatter \citep{wang2020xgass}. 
We can thus study the secondary dependence of the MZR on $\hi$ in a spatially more coherent way. We emphasize that our goal in this paper, as a first step, is merely to test, with the best sample available, whether the MZR shows stronger secondary dependence on the newly derived inner $\hi$ masses than the total $\hi$ masses and SFR. If so, this would strongly suggest that future observational and theoretical studies of the MZR should consider the distribution and radial flow of newly accreted gas. We do not intend to establish or calibrate a fundamental relation in the parameter space of metallicity, stellar mass, inner $\hi$ mass, $\hi$ mass, and SFR, which would require more stringent considerations of sample completeness and the separations of observational uncertainty and intrinsic scatter\citep{cresci2019fundamental}. 

The paper is structured as follows. In Section \ref{sec:2} we describe the samples considered in this paper and their properties. In Section \ref{sec:3} we show our main results. 
Our main conclusions and discussion are presented in Section \ref{sec:4}.

\section{Data} \label{sec:2}
\subsection{The HI-MaNGA sample}
Recently, the seventeenth data release of SDSS (DR17)  released integral field spectroscopy observations and ancillary data products of 10,010 galaxies from the MaNGA survey \citep{abdurrouf2022seventeenth}. It provides Data Analysis Pipeline products \citep{westfall2019data,law2021sdssiv}\footnote{\url{https://www.sdss.org/dr17/manga/manga-analysis-pipeline/}}, which give easy access to measurements of Balmer series lines and strong forbidden lines. We use the parameters measured from the MaNGA-Pipe3D value-added catalog (\citealt{lacerda2022pyfit3d}, Sánchez et al. in prep). We use the Primary $+$ MaNGA subsample for our analysis, as the Secondary sample lies at larger distances and has a 
lower $\hi$ detection rate. The sample has a flat stellar-mass distribution. 
The SDSS photometric measurements and estimated stellar masses are taken from the NASA-Sloan Atlas catalog \citep{blanton2007kcorrections}. 
We only select galaxies with stellar masses in the range $10^9<M_*/\Msun<10^{11}$, to limit the metallicity range and thus reduce the diagnostic uncertainties (see section \ref{sec:2.2}), as well as to enable a convenient comparison with the literature \citep{brown2018role}.
We select disk-like galaxies by requiring the $r$-band light concentration $r_{90}/r_{50}<2.7$, where $r_{50}$ and $r_{90}$ are the radii that enclose 50\% and 90\% of the total flux, respectively. Because the technique developed in \cite{wang2020xgass} to estimate the $\hi$ mass within the optical radius can only be applied to disk-like galaxies, we also require galaxies to have $b/a>0.5$, to exclude highly inclined galaxies. These selections result in 1684 galaxies.

To ensure a reliable estimate of the SFR and gas-phase metallicity based on calibrated strong-line diagnostics, we select SFGs by adopting the ionizing classification scheme of \cite{sanchez2021global}. We select galaxies that have EW($\ha$)$>6 \rm \AA$ and locate them below the \cite{kewley2001theoretical} demarcation line in the $\oiii/\hb$ versus $\nii/\ha$ Baldwin, Phillips, and Telervich \citep{baldwin1981classification} diagnostic diagram, where the line ratios are for the region within the central 2.$''$5  aperture. After excluding these galaxies, the sample size reduces to 1111. 

The recent third data release of \hi-MaNGA \citep{law2021sdssiv} provides 6358 unique galaxies, with $\hi$ data taken from Green Bank Telescope (GBT) observations and the Arecibo Legacy Fast ALFA (ALFALFA) survey \citep{haynes2018arecibo}.
The catalog provides flags as to whether a spectrum is potentially confused with emission from neighboring galaxies, by checking the possible overlapping in velocities and projected distances in comparison to 1.5$\times$HPBW (half-power beamwidth).
Following the procedure of \cite{stark2021imanga}, we clean the dataset by firstly excluding any confused spectra, and then selecting the spectrum with the better signal-to-noise ratio (S/N), if both GBT and ALFALFA spectra are available for one galaxy. 
After requiring $\hi$ data availability, we obtain 789 galaxies, of which 596 galaxies have an $\hi$ flux detected with an S/N$\geqslant$3. They constitute the final whole sample and the final $\hi$ detected sample, respectively, that are used in this study.

\subsection{The SFR and Gas-phase Metallicity}\label{sec:2.2}
The SFR and  gas-phase metallicity are taken from Pipe3d catalog. The integrated SFR is derived from the dust attenuation-corrected $\ha$ luminosity and the gas-phase metallicity is derived with strong-line methods. Because each strong-line method has its systematic limitations \citep{kewley2002using}, we use several methods that are commonly used. 
The strong-line methods considered in this paper include the N2O2 method \citep{kewley2002using}, the O3N2 method \citep{marino2013o3n2}, and the R23 method \citep{curti2020massmetallicity}. In the analysis, we use the O3N2 index as the fiducial estimate of the gas-phase metallicity. We present results based on the other two types of metallicity estimates in Section \ref{sec:other_calibrator}, and point out that the results are consistent. We mainly use the metallicity at 1$R_e$, which is derived as the average within an annular ring of 0.75-1.25 $R_e$, in the following analysis. But we also use the metallicity in the central region (2.$''$5  in radius) for a comparison.

\subsection{The Inner $\hi$ Mass}
We derive the inner $\hi$ mass within the optical $r_{90}$ following the method of \cite{wang2020xgass}.  We estimate the characteristic size of the $\hi$ disk $\rhi$ (defined as the semi-major axis of the 1 $M_{\odot}$ pc$^{-2}$ isophote)  based on the size--mass relation of $\hi$ \citep{swaters2002westerbork, wang2016new}. We take the median profile of the $\hi$ mass surface density as a function of the $\rhi$-normalized radius from \cite{wang2016new}, and use $\rhi$ to scale its radius. We then integrate the profile from $1.5 \rhi$ to the $r$-band $r_{90}$ to infer the $\hi$ mass beyond $r_{90}$ ($\mhiout$) if $1.5 \rhi>r_{90}$. The $\mhiout$ is taken to be zero if $1.5 \rhi<r_{90}$. Finally, $\mhiin= \mhi -\mhiout$. We refer the readers to \cite{wang2020xgass} for more details. 

As pointed out in \cite{wang2020xgass}, it is less reliable to derive $\mhiin$ within radii smaller than $r_{90}$, as individual galaxies start to deviate significantly from the normalized median profile of $\hi$ surface density. So $M_{\rm HI,in}$ are not derived for exactly the same region ($r\sim R_e$) where IFU measurements have been taken.

\subsection{Derivation from Main Sequences: $\Delta X$ } \label{sec:2.4}
In this study, we investigate the secondary dependencies of the MZR. In the following, we firstly define the MZR based on the MaNGA DR17 SFG sample, then derive the scatters based on it. 

\begin{figure}[htb]
\includegraphics[width=0.5\textwidth]{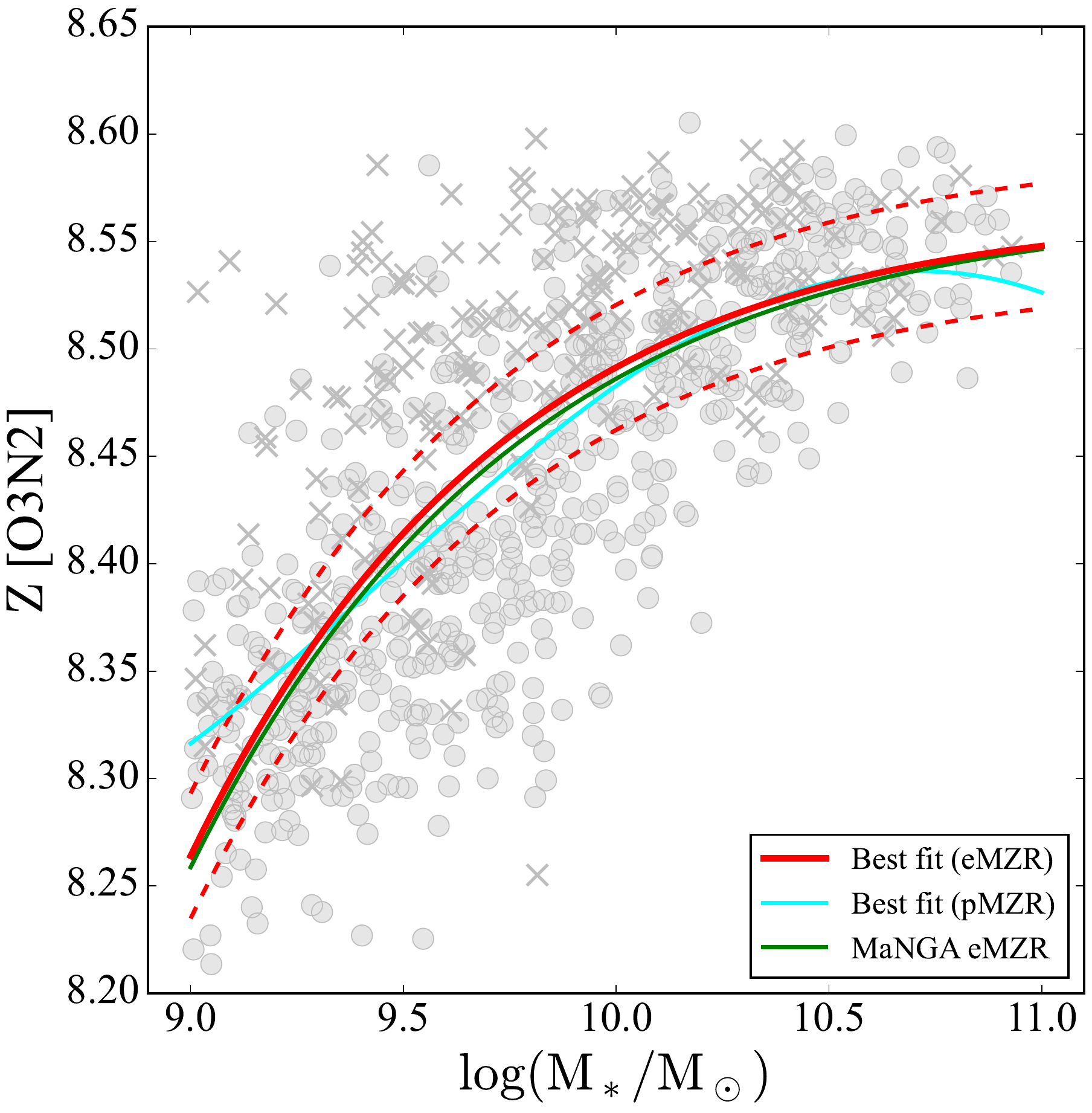}
\caption{The mass-metallicity relation. The red curve is the best-fit exponential relation of the MZR for the MaNGA DR17 SFG sample, and the red dashed curves are 0.5$\sigma$ either side of the best fit. The cyan curve is the fourth-order polynomial MZR. The green curve is the exponential MZR from \cite{barrera-ballesteros2017separate}. The circles show the $\hi$ detected galaxies, and the crosses show the $\hi$ undetected galaxies.}
\label{fig:mzr}
\end{figure}

In Figure \ref{fig:mzr}, we fit the MZR by using a fourth-order polynomial function and an exponential function, and compare the best fits with the MZR in \cite{barrera-ballesteros2017separate}. We can see that our best-fit exponential relation agrees very well with that of \cite{barrera-ballesteros2017separate}, but our best-fit exponential relation differs from our best-fit polynomial relation at the two ends. We thus use the exponential function in the following analysis, but confirm that changing the function formula does not significantly change the main results in the following. We derive the MZR by ourselves, instead of directly using the one in \cite{barrera-ballesteros2017separate}, because not all the metallicity indexes (e.g. the N2O2 metallicity) used in this paper have an MZR in \cite{barrera-ballesteros2017separate}. 

\begin{figure*}[htb]
\includegraphics[width=\textwidth]{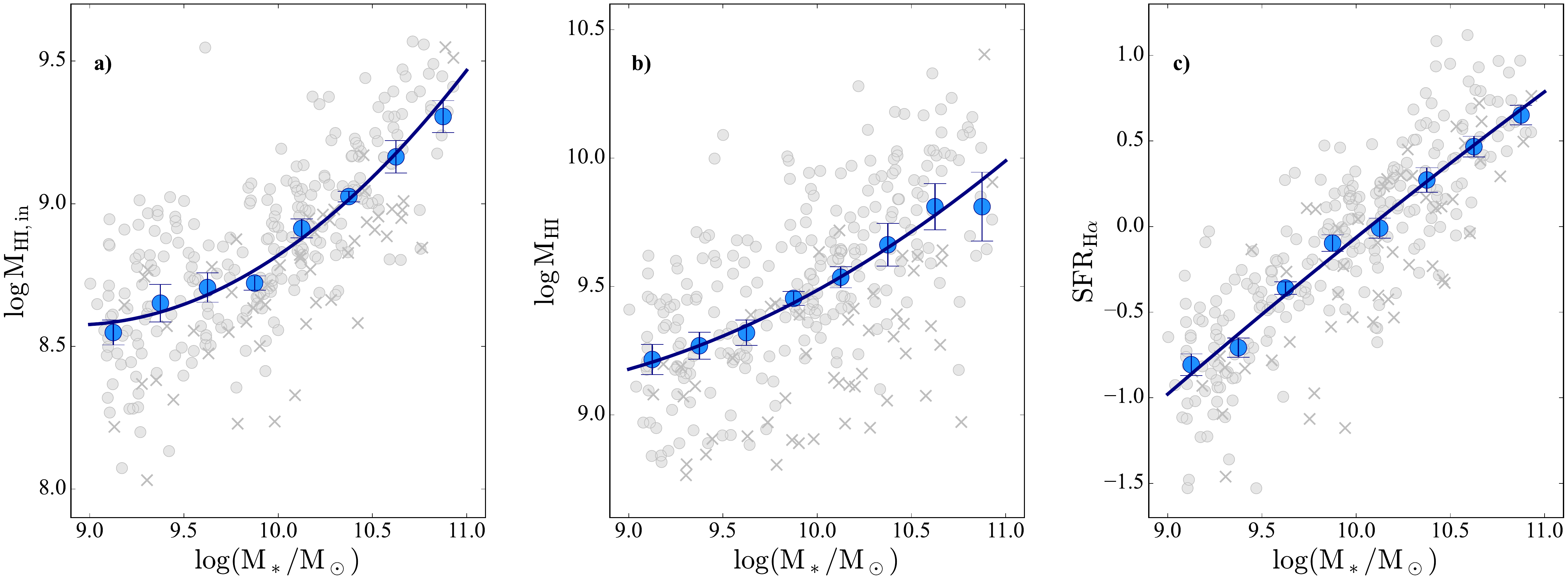}
\caption{Reference relations for deriving the deviation of the galaxies from the population near the MZR. In each panel, the circles show the $\hi$ detected reference sample, and the crosses show the $\hi$ undetected reference sample. The large blue circles indicate the medians of the parameter in $M_*$ bins, with errors derived through bootstrapping, and the blue curves show the best-fit second-order polynomial relations to the large blue circles.
}
\label{fig:mzr2}
\end{figure*}

We select the galaxies within $\pm 0.5\sigma$ of the MZR, and use them as the reference sample for deriving the reference scaling relations. When deriving the reference $M_{\rm HI}$ relation, we firstly divide the reference sample into eight $M_*$ bins and obtain the median $M_{\rm HI}$ in each $M_*$ bin. We then fit a second-order polynomial equation to the relation of the median $M_{\rm HI}$ as a function of the $M_*$ bin. We do the same for $M_{\rm HI,in}$ and the SFR. These median relations and best-fit polynomial relations are displayed in 
Figure~\ref{fig:mzr2}. We emphasize that these are median relations for the reference sample, rather than for general galaxies. 
Below, we provide the best-fit equation for the relation between $\mhiin$ and $M_*$:
\begin{equation}
    {M}_{\rm HI,in}  =  0.1356\times \log(M_*)^2  - 2.2823 \times \log(M_*) + 18.1015
\end{equation}

Then, for each galaxy in our sample, we calculate its vertical distance from the reference scaling relation of parameter $X$ as $\Delta X$, where $X$ can be $M_{\rm HI}$, $M_{\rm HI,in}$ and SFR. The $\Delta X$ thus quantifies the deviation of a galaxy from the population close to the MZR in the space of $X$ at a given $M_*$.
We also tried an alternative method of deriving $\Delta X$, by selecting a control sample with similar stellar masses for each galaxy from the reference sample, and calculating the difference of the galaxy from the median value of the control sample. The trends presented in Section \ref{sec:3} did not significantly change.

\begin{figure*}[htb]
\includegraphics[width=\textwidth]{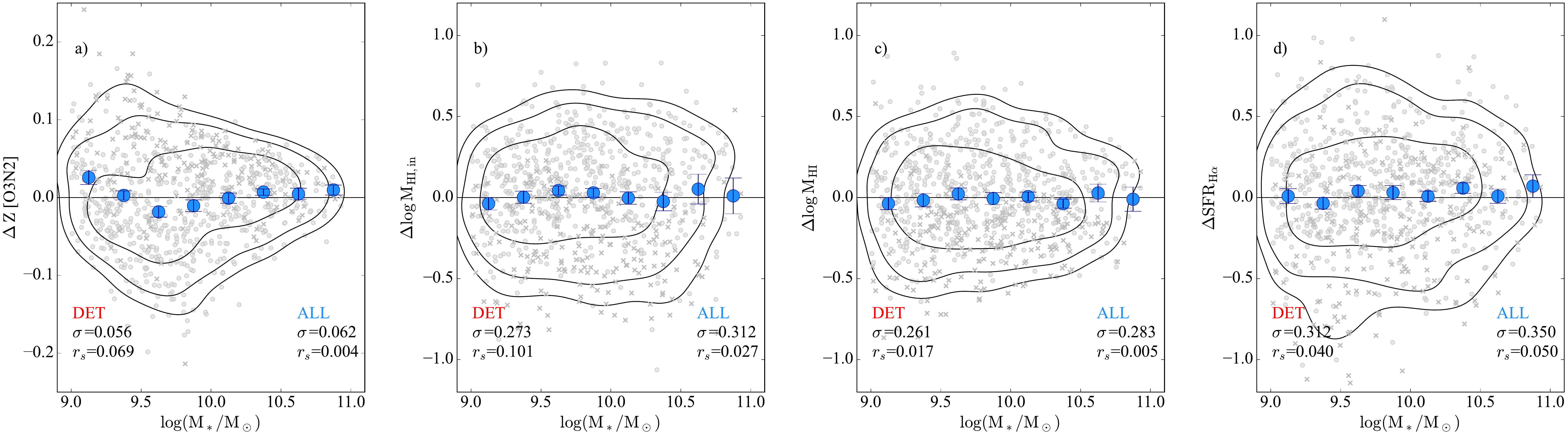}
\caption{The dependence of $\Delta Z$ and $\Delta X$ on $M_*$. From left to right, X is defined as $\mhi$, $\mhiin$, $\rm SFR$. 
In each panel, the circles show the $\hi$ detected galaxies, and the crosses show the $\hi$ undetected  galaxies. The large blue circles indicate the medians of parameters for the whole sample using the upper limits for nondetections in each $M_*$ bin, with the errors derived through bootstrapping. The contours represent the density distribution, with each contour encircling 90\%, 80\%, and 50\% of the points, respectively.
The Spearman correlation coefficient ($r_s$) and standard deviation error are given for detections only (bottom left corners) and for the whole sample, using upper limits for nondetections (bottom right corners).}
\label{fig:x-m}
\end{figure*}

We check the residuals from the MZR relation and reference scaling relations derived above in Figure \ref{fig:x-m}. We use the Spearman correlation coefficient $r_s$, and we use a threshold of 0.4 for correlation coefficients to indicate moderate correlations.
In Figure 3(a), we show that the correlation of the residuals from the MZR relation ($\Delta Z$) with $M_*$ is pretty weak.
The standard deviation of the residuals of the metallicity with respect to the best-fit MZR is 0.062, and the correlation coefficient is 0.004. It should be noticed that when we fit the MZR for all MaNGA SFGs, the scatter is 0.057. 
In panels (b), (c), and (d), we show $\dmhiin$, $\dmhi$, and $\dsfr$ as a function of $M_*$, and the correlation coefficients are all low (0.027, 0.005, and 0.050, respectively). So we conclude that the residuals from the relations do not retain any significant residual correlations with the stellar mass.

\section{Results} \label{sec:3}
\subsection{MZR Scatter Dependent on Inner $\hi$ Mass}
\label{sec:3.1}
\begin{figure*}[htb]
\includegraphics[width=\textwidth]{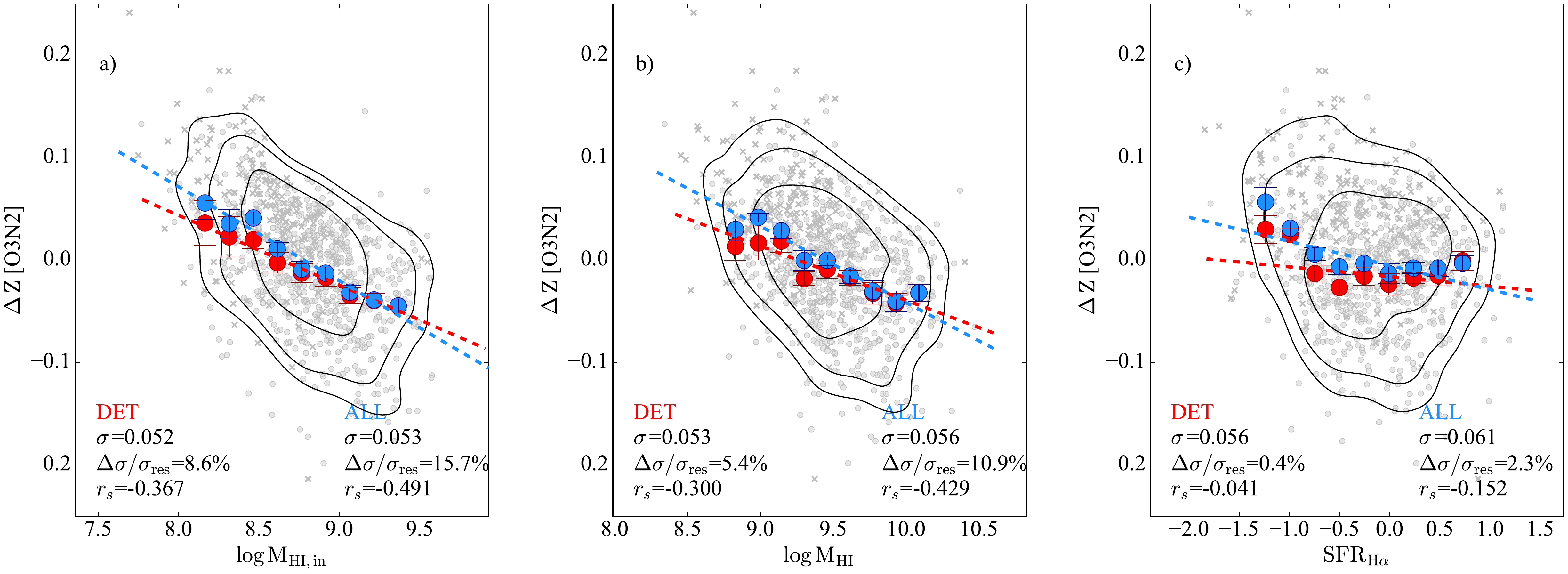}
\caption{The relation between  of $\dz$ and X.  From left to right, X is defined as $\mhiin$, $\mhi$, and $\rm SFR$, respectively. 
In each panel, the circles show the $\hi$ detected galaxies, and the crosses show the $\hi$ undetected  galaxies. The large blue circles indicate the  medians of $\dz$ in the $M_*$ bins with errors derived through bootstrapping, and the blue curve are the best-fit relation to the large blue circles. Contours represent the density distribution, with each contour encircling 90\%, 80\%, and 50\% of the points, respectively. 
The large red circles indicate the medians of $\dz$ for the $\hi$ detected subsample, and the red curve is the best-fit relation for the large red circles. 
The Spearman correlation coefficient ($r_s$), scatter reduction, and standard deviation error are given for detections only (bottom left corners) and for the whole sample, using upper limits for nondetections (bottom right corners).}
\label{fig:dz-x}
\end{figure*}

Here we compare the relations of $\dz$ versus $\mhiin$, $\mhi$, and SFR. In the following, we mainly discuss statistics based on the whole sample, including galaxies undetected in $\hi$, but the statistics based on the $\hi$ detected sample are also present for a comparison. We point out that the results are consistent. 
In Figure \ref{fig:dz-x}, we show that $\dz$ has a  moderate correlation with $\mhiin$ and $\mhi$, and that the correlation coefficient of $\mhiin$ is larger than the correlation coefficient of $\mhi$. On the contrary, the correlation coefficient of SFR is much weaker than the $\hi$-related  relations. 
Following \cite{alvarez-hurtado2022which},
we perform linear regression for the relation of $\dz$ versus $X$. The scatters of the linear regressions, $\sigma_{res}$, are  0.053, 0.056, and 0.061, respectively, for $\mhiin$, $\mhi$, and SFR. We compare these scatters with the scatter of the MZR $\sigma_{\rm \Delta MZR}$, and calculate $\rm \Delta\sigma/\sigma_{res}$, where $\Delta\sigma = (\sigma_{\rm \Delta MZR} -\sigma_{res} )$. 
When considering a secondary dependence, the scatter of the MZR is reduced by 15.7\% for $\mhiin$ and 10.9\% for $\mhi$, much more than for SFR (2.3\%). The very small reduction in scatter when considering a secondary dependence on SFR has been noticed in the literature  \citep{barrera-ballesteros2017separate,sanchez2017massmetallicity,sanchez2019sami,alvarez-hurtado2022which}.
We notice that the reduction in scatter by including $\mhi$ is much smaller in \cite{alvarez-hurtado2022which} (by 3\%) than it is in our analysis. Some possible reasons are that the $\hi$ data here are more homogeneous (mostly observed by GBT and ALFALFA) and complete (nondetections are included as upper limits), and the number of galaxies detected in $\hi$ is larger (596 in comparison to 167 in \citealt{alvarez-hurtado2022which}) in our analysis. 
We emphasize that the major result to highlight here is that $\mhiin$ indeed seems to be better related to $\Delta Z$ than $\mhi$.

\begin{figure*}[htb]
\includegraphics[width=\textwidth]{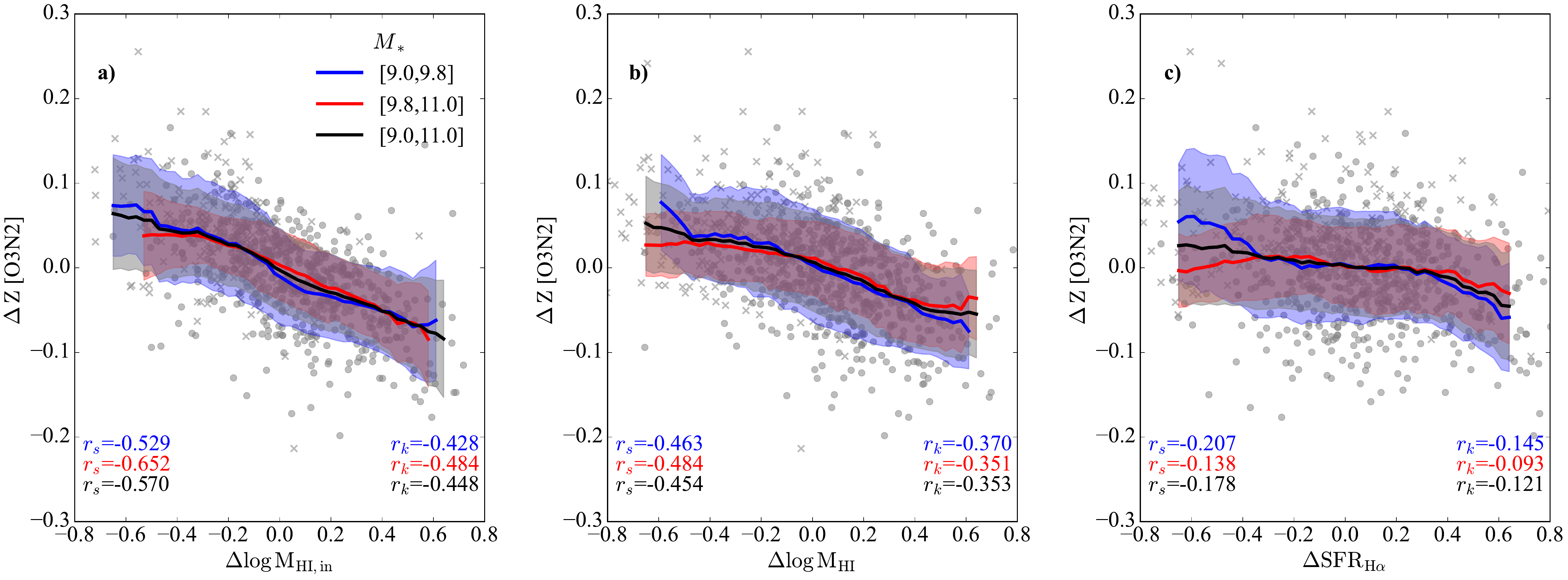}
\caption{
The relation between $\Delta Z$ and $\Delta X$. From left to right, X is defined as $\mhiin$, $\mhi$, and $\rm SFR$. The circles show the $\hi$ detected galaxies, and the crosses show the $\hi$ undetected  galaxies. The black solid curves are the median relations, and the widths of the shaded regions represent the 1$\sigma$ scatters of the distribution. The blue and red curves and shades are similar to the black ones, but for subsample of $9<\log M_*/\Msun<9.8$ and $9.8<\log M_*/\Msun<11$.  The text in the bottom left and right corner shows the Spearman and Kendall correlation coefficients of $\Delta Z$ and $\Delta X$ for the $\hi$ detected sample and the whole sample, respectively.
}
\label{fig:dxdy}
\end{figure*}

We further study the relations of $\dz$ versus $\dmhiin$, $\dmhi$, and $\dsfr$, to remove the dependence on $M_*$ from both $Z$ and the interested parameters. We use the Spearman correlation coefficient $r_s$,  the Kendall rank correlation coefficient $r_k$, and the best-fit linear slope to quantify the correlation strength. The Spearman coefficients are only derived for the $\hi$ detected galaxies, and the Kendall rank $r_k$ and linear slopes are derived for the whole sample of galaxies, including both those detected and undetected in $\hi$. The upper limits of the $\hi$ masses are used for the $\hi$ undetected galaxies, and the astronomical survival analysis \citep{isobe1986statistical} is conducted when upper limits are involved. 
We summarize these derived statistical values in Table \ref{tab:1}, and show the trend in Figure \ref{fig:dxdy}. 

\begin{splitdeluxetable*}{ccccccccBccc}
\tablecaption{The Correlation Coefficients and Linear Slope for the Relation between $\Delta Z$ and $\Delta X$}
\tablehead{
\colhead{Metallicity} & \colhead{$\Delta$X} & \multicolumn3c{$9<\log M_*/\Msun<9.8$} &  \multicolumn3c{$9.8<\log M_*/\Msun<11$} & \multicolumn3c{$9<\log M_*/\Msun<11$} \\
\colhead{} & \colhead{} & \colhead{Spearman $r_s$} & \colhead{Kendall $r_k$} & \colhead{Slope} &   \colhead{Spearman $r_s$ } & \colhead{Kendall $r_k$} & \colhead{Slope} &  \colhead{Spearman $r_s$} & \colhead{Kendall $r_k$} & \colhead{Slope}  }
\startdata
O3N2 &\mhiin &-0.529(3.6e-23)&-0.428(3.6e-37)&-0.123(0.003)& -0.652(6.4e-37)&-0.484(2.0e-46)&-0.106(0.006)&-0.570(1.0e-52)&-0.448(2.7e-79)&-0.119(0.002)\\
O3N2 &\mhi &-0.463(1.9e-17)&-0.370(3.6e-28)&-0.118(0.003)& -0.484(1.2e-18)&-0.351(3.1e-25)&-0.069(0.003)&-0.454(1.1e-31)&-0.353(6.1e-50)&-0.090(0.002)\\
O3N2 &${\rm SFR}$ &-0.207(3.0e-04)&-0.145(1.7e-04)&-0.073(0.004)& -0.138(1.8e-02)&-0.093(1.8e-02)&-0.021(0.003)&-0.178(1.3e-05)&-0.121(9.6e-06)&-0.046(0.002)\\
O3N2 (center) &\mhiin &-0.267(2.6e-06)&-0.262(6.2e-15)&-0.087(0.003)& -0.173(3.0e-03)&-0.162(1.6e-06)&-0.044(0.007)&-0.229(1.6e-08)&-0.224(3.8e-21)&-0.070(0.003)\\
O3N2 (center) &\mhi &-0.279(8.2e-07)&-0.247(2.1e-13)&-0.082(0.003)& -0.167(4.0e-03)&-0.157(3.4e-06)&-0.029(0.002)&-0.214(1.3e-07)&-0.203(1.5e-17)&-0.052(0.002)\\
O3N2 (center) &${\rm SFR}$ &-0.311(3.3e-08)&-0.218(1.7e-08)&-0.093(0.005)& -0.148(1.1e-02)&-0.098(1.2e-02)&-0.024(0.003)&-0.220(6.1e-08)&-0.151(3.7e-08)&-0.057(0.003)\\
\enddata
\label{tab:1}
\tablecomments{The statistics are derived for the low-mass and high-mass sub-samples and for the whole sample. They are also derived for $\Delta Z$ derived in center region and  in $\sim R_e$  regions. The values in brackets after correlations coefficients are the corresponding p-values, and those after the slopes are 1$\sigma$ error.}
\end{splitdeluxetable*}

From Figure \ref{fig:dxdy}, we can see that $\dz$ is anticorrelated with $\dmhiin$ and $\dmhi$, but only shows a weak correlation with $\dsfr$. 
The $\dmhi$ relation is stronger than that of $\dsfr$, consistent with the past findings of \cite{brown2018role}.
But $\dmhiin$ shows the strongest anticorrelation. 
The slope of the relation involving $\dmhiin$ is larger than that involving $\dmhi$, especially for the larger mass bin. 
These results strongly support our speculation that metal evolution is local, and that the quantities involved in the investigation of the MZR should be derived from similar regions. 

We present below the best-fit linear relation for the black curve in panel (a) of Figure \ref{fig:dxdy}:
\begin{equation}
    \Delta  Z =  -0.119 \times \Delta \log {\rm M_{HI,in}}  - 0.003 
\end{equation}

\subsection{A Test for Artificial Correlations Caused by Measurement Uncertainties in the $\Delta$ Relations}
\begin{figure*}[htb]
\includegraphics[width=\textwidth]{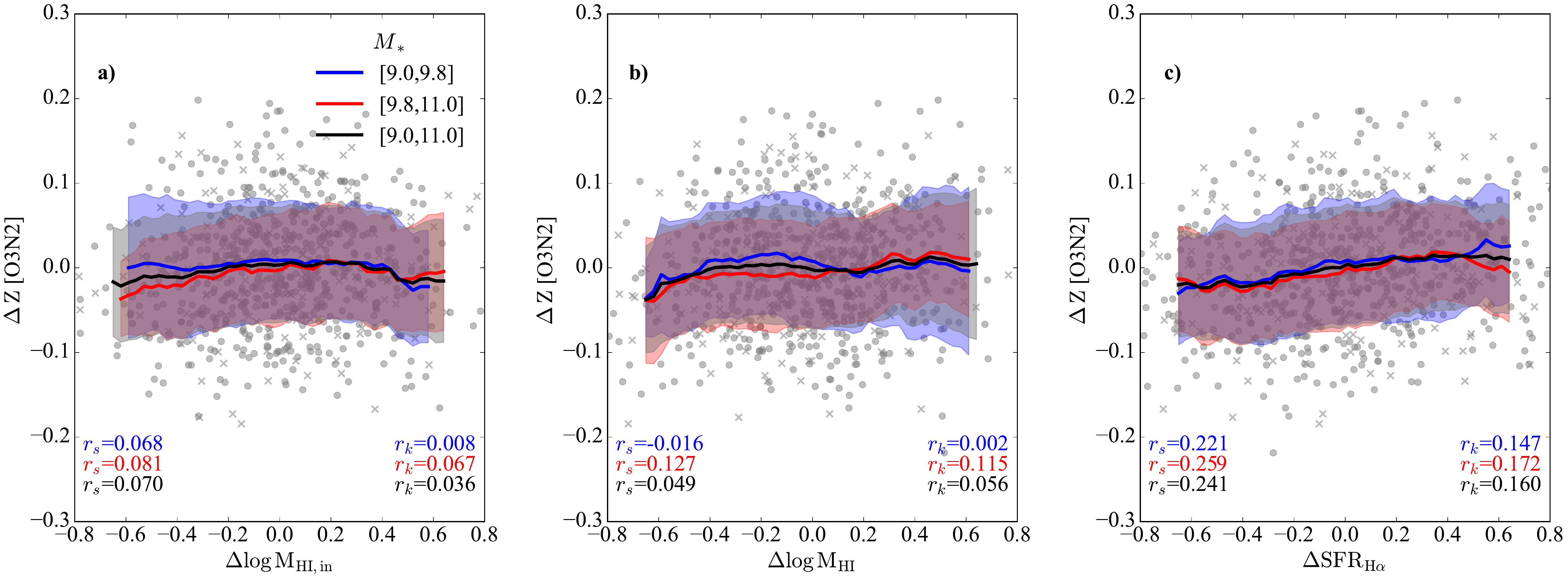}
\caption{A test for the possibility of artificial relations due to error propagation. The $\Delta$ relations are similar to those in Figure~\ref{fig:dxdy}, but the dataset has been simulated according to the reference best-fit relations  and thescatter of those relations. Please see the text for details. }
\label{fig:sim}
\end{figure*}

\cite{sanchez2021edgecalifa} have used EDGE--CALIFA survey data to show that the resolved properties, including $\Sigma_{\rm SFR}$, $\Sigma_*$, and $\Sigma_{\rm mol}$, show a set of power-law relations between each other, but that the scatters around those relations are dominated by measurement errors, which later cause artificial $\Delta$ relations. 
Here, in this section, we show a simple test to check the reliability of our correlation and fitting results, following the procedure introduced in \cite{sanchez2021edgecalifa}. We  simulate a random distribution along the best-fit relations of $Z$, $\mhi$, $\mhiin$, and $\rm SFR$ as a function of $M_*$. Taking the simulated relation of $Z$ versus $M_*$ as an example: for a given $M_*$ we derive its related $Z$ based on the best-fit $Z$ versus $M_*$ relation, then we randomly shift the $Z$ value by extent, following a Gaussian distribution with $\sigma$ equivalent to the standard deviation of the relation (i.e. the standard deviation of $\Delta Z$). We do the same for $\mhi$, $\mhiin$, and $\rm SFR$, using the reference relations that we derived above. Additionally, we randomly shift the $M_*$ values by extent, following a Gaussian distribution with $\sigma$ equivalent to the typical measurement error of 0.2 dex. We thus produce a simulated sample of galaxies with sizes equivalent to our analysis sample. We then investigate whether  we can still find the relations revealed in Figure~\ref{fig:dxdy} with this simulated data set. 
In Figure~\ref{fig:sim}, we show one example of the simulation results, which we repeat 1000 times, with the median value of  Kendall rank  $r_k$ for $\dmhiin$, $\dmhi$, and $\dsfr$ being 0.049, 0.050 and 0.116, respectively.  As we can see, there are no significant correlations between these simulated parameters, in contrast to what we found in Figure~\ref{fig:dxdy}. This result indicates that the trends of the $\hi$-related parameters present in Figure \ref{fig:dxdy} are not an artifact of the propagation of errors through primary relations with the stellar mass.  On the other hand, the correlation for SFR in the simulation test is of the same order as was found in Figure~\ref{fig:dxdy}, indicating that dependence of the MZR scatter on SFR is really weak or negligible.

\subsection{Further Dependence on the Radial extension}
In order to further demonstrate that the correlation between the metallicity and $\hi$ gas mass weakens if there is a large difference in radial extension between the two quantities, we check the correlation of the metallicity in the central region (2.$''$5 in radius) with $\mhiin$, $\mhi$ and SFR. Among our sample, 98.0\% of the galaxies have $R_e>2.''5$.
We repeat the analysis of Section \ref{sec:3.1}, with the metallicity at $\sim R_e$ replaced by the metallicity in the center.

\begin{figure*}[htb]
\includegraphics[width=\textwidth]{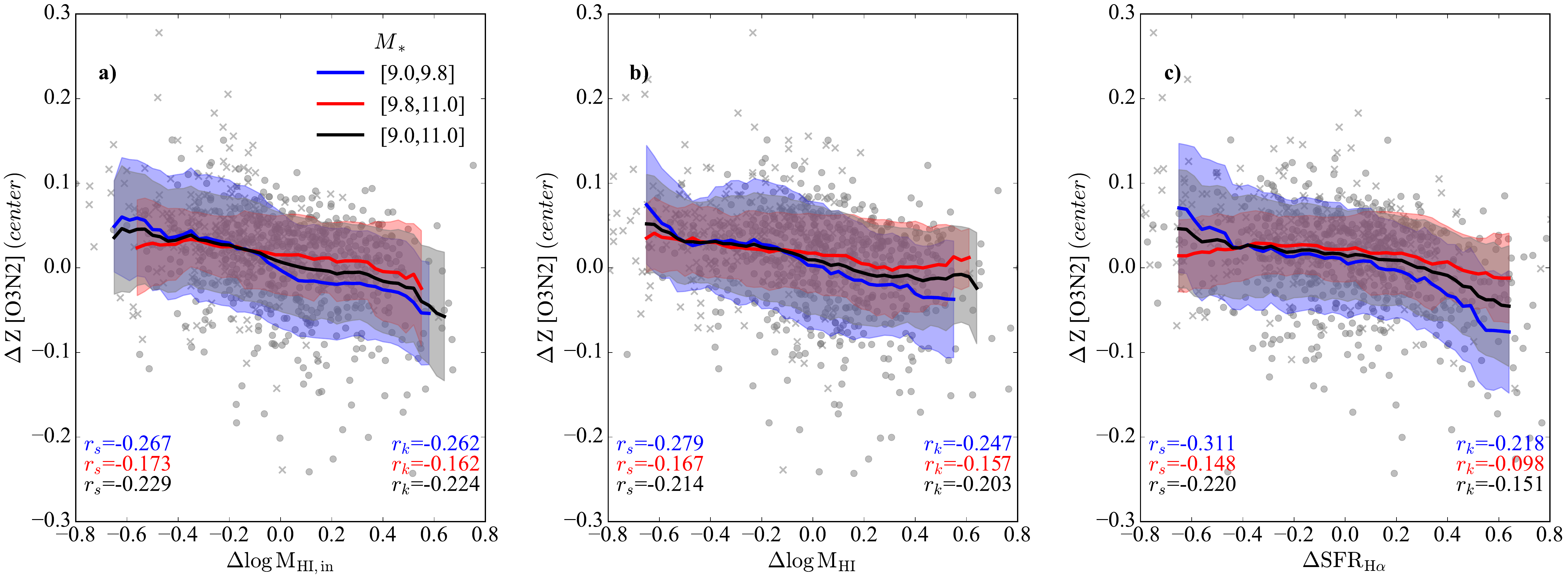}
\caption{The relation between $\Delta Z$ and $\Delta X$ in the center regions. Similar to Figure ~\ref{fig:dxdy}, but the $Z$ at $\sim R_e$ is replaced by the median $Z$ derived in the center regions.
}
\label{fig:radius_dxdy}
\end{figure*}

The results are presented in Figure \ref{fig:radius_dxdy}.
We find that $\dmhiin$ is indeed less correlated with the $\Delta Z$ in the central region than with the $\Delta Z$ at $\sim R_e$, with a Spearman correlation coefficients of -0.229 for the former and -0.570 for the latter.
And the slope for the relation in the center is also much shallower than that for $\sim R_e$. This confirms our speculation that when sampling from different radial ranges, the relation between the metallicity and $\hi$ gas weakens. 

As for the SFR, we find a slight enhancement of the correlation between $\dz$ and $\dsfr$ when using the central metallicity. This slight enhancement may be caused by the complex situation of the data quality and ionization in the central region, including stronger contamination
by diffuse ionized gas, a lower S/N ratio and lower contrast between the emission lines, possible contamination by weak/strong active galactic nuclei, and/or shocks due to central galactic winds. The Spearman correlation coefficient is still low, despite the sight increase, thus it cannot support the existence of a significant secondary dependence of the MZR scatter on SFR, consistent with the findings in the literature \citep[e.g.][]{sanchez2017massmetallicity,sanchez2019sami}.

\subsection{The Results for the N2O2 Index and the R23 Index}
\label{sec:other_calibrator}
\begin{figure*}[htb]
\includegraphics[width=\textwidth]{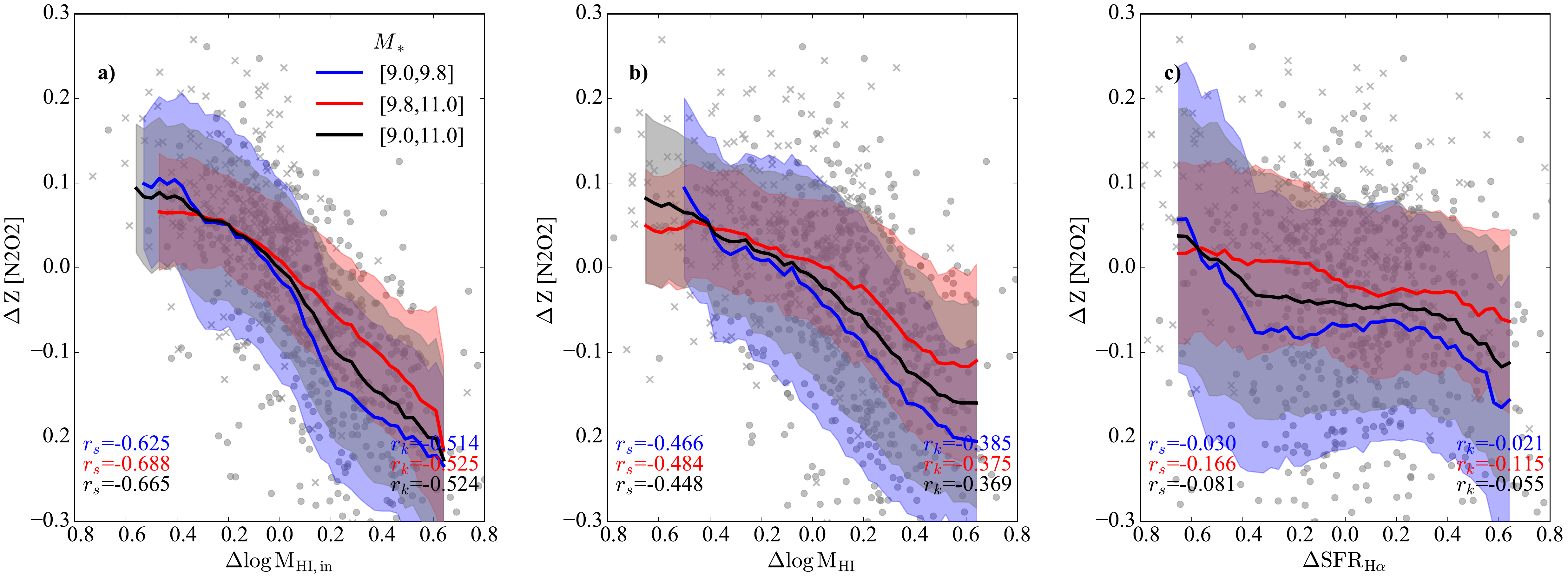}
\includegraphics[width=\textwidth]{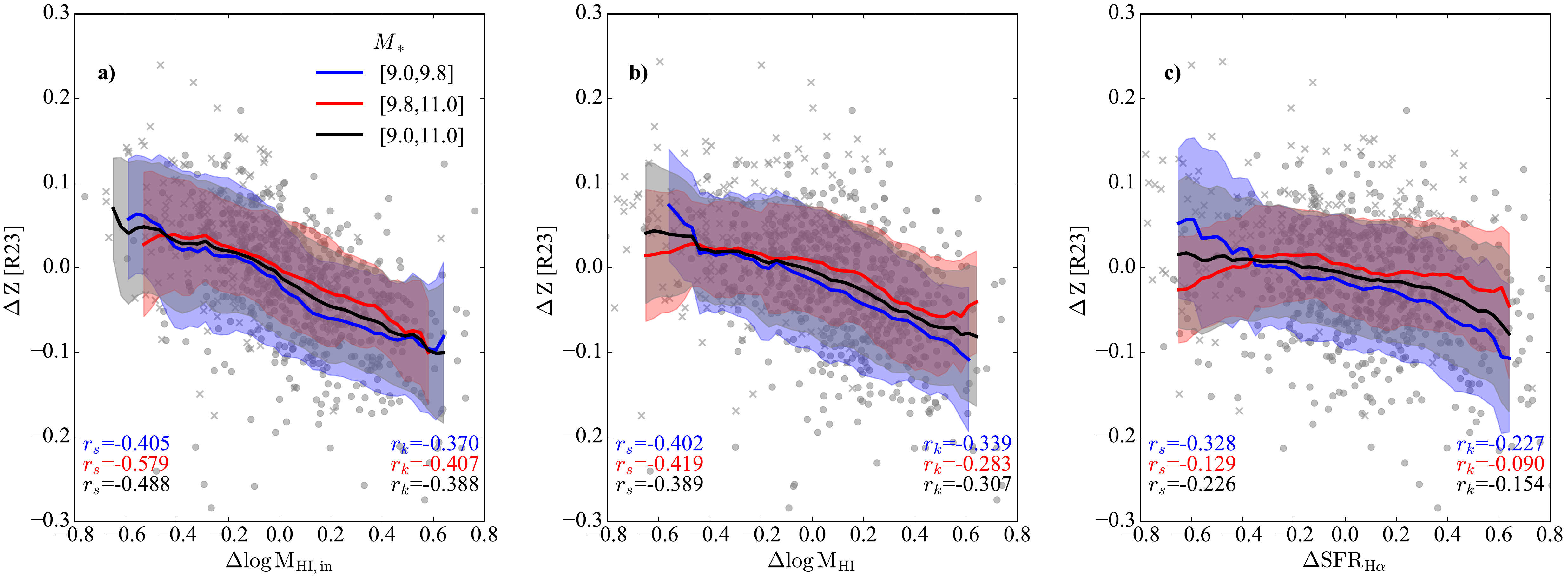}
\caption{The relation between $\Delta Z$ and $\Delta X$ for the N2O2 (upper) and R23 (bottom) indexes. Similar to Figure ~\ref{fig:dxdy}, but the median $Z$ is derived using the N2O2 and R23 indexes. 
}
\label{fig:a1}
\end{figure*}

In addition, we present the results for the N2O2 index and the R23 index in Figure \ref{fig:a1}, and we summarize these derived statistical values in Table \ref{tab:A}.
We find that the anticorrelation of $\mhiin$ with the scatter of the MZR is robust when using different diagnostics and different mass bins. The Spearman correlation coefficients range between -0.405 and -0.688. For $\mhi$, the correlations are slightly weaker, with the Spearman correlation coefficients ranging between -0.389 and -0.484. 

On the other hand, the correlation strength of the SFR with the scatter of the MZR can vary between being weak and being negligible when we change the diagnostics and mass ranges. This indicates that the secondary dependence of the MZR on the SFR may be really weak and depend on the calibrator, which has been reported before in the literature \citep[e.g.][]{kashino2016hideandseek,telford2016exploring}.

\begin{splitdeluxetable*}{ccccccccBccc}
\tablecaption{The Correlation Coefficients and Linear Slope for the Relation between $\Delta Z$ and $\Delta X$ for the N2O2 and R23 Indexes}
\tablewidth{0pt}
\tablehead{
\colhead{Metallicity} & \colhead{$\Delta$X} & \multicolumn3c{$9<\log M_*/\Msun<9.8$} &  \multicolumn3c{$9.8<\log M_*/\Msun<11$} & \multicolumn3c{$9<\log M_*/\Msun<11$} \\
\colhead{} & \colhead{} & \colhead{Spearman $r_s$} & \colhead{Kendall $r_k$} & \colhead{Slope} &   \colhead{Spearman $r_s$ } & \colhead{Kendall $r_k$} & \colhead{Slope} &  \colhead{Spearman $r_s$} & \colhead{Kendall $r_k$} & \colhead{Slope}  }

\startdata
N2O2 &\mhiin &-0.625(3.5e-34)&-0.514(7.2e-53)&-0.324(0.008)& -0.688(1.6e-42)&-0.525(2.9e-54)&-0.210(0.010)&-0.665(2.2e-77)&-0.524(9.6e-108)&-0.283(0.007)\\
N2O2 &\mhi &-0.466(1.2e-17)&-0.385(2.1e-30)&-0.273(0.005)& -0.484(1.1e-18)&-0.375(1.5e-28)&-0.148(0.007)&-0.448(9.9e-31)&-0.369(3.5e-54)&-0.206(0.005)\\
N2O2 &${\rm SFR}$ &-0.030(6.0e-01)&-0.021(5.8e-01)&-0.105(0.011)& -0.166(4.2e-03)&-0.115(3.2e-03)&-0.062(0.003)&-0.081(4.7e-02)&-0.055(4.4e-02)&-0.086(0.005)\\
R23 &\mhiin &-0.405(2.5e-13)&-0.370(3.9e-28)&-0.138(0.003)& -0.579(1.0e-27)&-0.407(2.3e-33)&-0.109(0.007)&-0.488(4.8e-37)&-0.388(6.3e-60)&-0.128(0.003)\\
R23 &\mhi &-0.402(3.8e-13)&-0.339(6.8e-24)&-0.140(0.004)& -0.419(6.4e-14)&-0.283(5.8e-17)&-0.070(0.005)&-0.389(5.1e-23)&-0.307(4.3e-38)&-0.101(0.002)\\
R23 &${\rm SFR}$ &-0.328(5.0e-09)&-0.227(4.2e-09)&-0.103(0.004)& -0.129(2.7e-02)&-0.090(2.2e-02)&-0.016(0.005)&-0.226(2.6e-08)&-0.154(1.8e-08)&-0.059(0.003)\\
\enddata
\label{tab:A}
\tablecomments{Similar to Table \ref{tab:1}, but for median $Z$ dervied using the N2O2 and R23 indexes and integrated $Z$ for all indexes.}
\end{splitdeluxetable*}

\section{Summary, Discussion, and Conclusion} \label{sec:4}

We use a disk-dominated subset of the \hi-MaNGA sample to study the secondary dependence of the MZR on $\hi$ properties. We derive the deviations of the galaxies from the galaxy population close to the MZR in the spaces of inner $\hi$ mass, $\hi$ mass, and SFR ($\dmhiin$, $\mhi$, and $\dsfr$ respectively).
Our main results are as follows: 
\begin{enumerate}
    \item Introducing a secondary relation with the inner $\hi$ mass can produce a small but noticeable decrease (16\%) in the scatter of MZR. And $\dz$ is more strongly correlated with $\dmhiin$ than with $\dmhi$. This result is consistent across different metallicity diagnostics and independent of stellar mass.

    \item We confirm the anticorrelation between $\dmhi$ and $\dz$,  but $\dsfr$ only shows a negligible or very weak correlation with $\dz$, which does not support the existence of a significant secondary dependence of the MZR scatter on the SFR.
    
    \item  $\dmhiin$ is more correlated with the $\dz$ derived at the $\sim R_e$ radius than at the center radius. 
    
\end{enumerate}

Theoretically, the evolution of gas-phase metallicity is driven by the local effects of physical processes, including dilution by gas accretion, mixing through gas transportation, the consumption of metals by SF, enrichment from evolved stars, and the loss of metals due to outflows \citep[e.g.][]{sharda2021physics}. Most of the oxygen was produced in the early periods of the evolutions of the galaxies, especially massive galaxies, and a significant episode of gas infall may also dilute the gas-phase metallicity.  The $\hi$ mass should thus be derived from roughly the same region as the metallicity, when its relation to the scatter of the MZR is investigated. 

Due to sensitivity limits, $Z$ is typically obtained well within the stellar disks. If the integral $\hi$ mass is used in studying the MZR, there is a risk of including the outlying $\hi$ \citep{swaters2002westerbork}, which typically has a metallicity lower than that of the stellar disks \citep{moran2012galex,carton2015gasphase}. 

The $\mhiin$ estimated within the optical $r_{90}$ in this paper attempt to mitigate the spatial inconsistency. They are still not perfectly estimated from exactly the same region as $Z$, but are much improved compared to the integral $\hi$ mass, in terms of having a closer radial range to $Z$. We emphasize such an improvement, as there is a large variation in the \hi-to-optical disk size ratio, with the predicted $R_{\rm HI}/R_{90}$ ranging from 1.644 to 6.452 (the 5th to 95th percentiles) in our $\hi$ detected sample (see also\citealt{wang2017local}). As we have shown, the $\dmhiin$ are much better anticorrelated with $\dz$ than $\dmhi$, consistent with $Z$ evolution being a local process, and strongly supporting our approach of using the $\mhiin$. Additional support for $Z$ being more closely linked with cospatial neutral gas comes from the result that the relation becomes significantly weaker when the $\dz$ values at $\sim R_e$ are replaced with the $\dz$ values in the central region. This is because the $\sim R_e$ region is closer to where the $\mhiin$ are derived (within $R_{90}$), where the neutral gas is more dominated by $\hi$ instead of by molecular gas \citep{leroy2008star}, and where the $\hi$ tends to be more pristine, instead of recycled \citep{mo1998formation}, as in the central region of galaxies. 

It is interesting to point out that although it is $\mhiin$ instead of $\mhi$ that directly dilutes $Z$, the large portion of $\hi$ stored beyond the optical radius is likely to continuously radially flow in \citep[e.g.][]{krumholz2018unified}, thus sustaining the diluting effect of $\mhiin$ before $\mhi$ is significantly depleted (on timescales of $>4$ Gyr; \citealt{saintonge2017xcold}). This dilution process will last for a long time and continuously affect the metallicity, which is different from the fast enrichment of stellar evolution \citep{maiolino2019re}. Galaxies with a higher integral $\hi$ mass fraction tend to have more $\hi$ within their stellar disks \citep{wang2020xgass}. This implies a quasi-equilibrium state of gas fueling in typical SFGs at low redshift, which may have partly caused the (weaker) anticorrelation between $\dz$ and $\dmhi$. 

We notice that the strongest anticorrelation and the steepest slope of $\dmhiin$ versus $\Delta Z$ hold for both low-mass and high-mass galaxies. But we also find that in the enhancement in correlation strength and the slope when using $\dmhiin$ instead of $\dmhi$ is more significant for high-mass galaxies than for low-mass galaxies. 
A possible reason is that the less massive gas-rich galaxies return to equilibrium more slowly \citep{lagos2016fundamental}, causing a larger scatter around the MZR. The enhancement in the correlation strength and the slope is weakened for low-mass galaxies, due to this larger scatter.

We remind the reader that our analysis is limited to disk-dominated galaxies, so future extension to more general samples with $\mhiin$ derived from real $\hi$ images will be useful. The statistics could also be improved by deeper surveys of $\hi$, with the upper limits replaced by real measurements. Despite these spaces for improvement, we conclude that our results strongly support $\hi$ properties being derived from the same region as $Z$ when studying the role of gas in regulating the MZR, as metal dilution and enrichment should be local processes. With relatively simple selection effects \citep{masters2019imanga, stark2021imanga}, the new relation between $\dmhiin$ and $\dz$ will hopefully provide a new input for constraining the metallicity evolution of galaxy models.


\section*{Acknowledgements}
We thank the anonymous referee for constructive comments. We thank all the people for useful discussions.
This work is supported by the Strategic Priority Research Program of the Chinese Academy of Sciences (No. XDB 41000000), the National Key R$\&$D Program of China (2017YFA0402600), the NSFC grant (Nos. 11973038,12073002,11721303), and the science research grants from the China Manned Space Project (Nos. CMS-CSST-2021-A07, CMS-CSST-2021-B02).

Funding for SDSS-IV has been provided by the Alfred P. Sloan Foundation and Participating Institutions. Additional funding toward SDSS-IV has been provided by the US Department of Energy Office of Science. SDSS-IV acknowledges support and resources from the Centre for High-Performance Computing at the University of Utah. The SDSS website is www.sdss.org.

SDSS-IV is managed by the Astrophysical Research Consortium for the Participating Institutions of the SDSS Collaboration including the Brazilian Participation Group, the Carnegie Institution for Science, Carnegie Mellon University, the Chilean Participation Group, the French Participation Group, Harvard-Smithsonian Center for Astrophysics, Instituto de Astrofísica de Canarias, The Johns Hopkins University, Kavli Institute for the Physics and Mathematics of the Universe (IPMU)/University of Tokyo, Lawrence Berkeley National Laborato y, Leibniz Institut fur Astrophysik Potsdam (AIP), Max-Planck-Institut fur Astronomie (MPIA Heidelberg), Max-Planck-Institut fur Astrophysik (MPA Garching), Max-Planck-Institut fur Extraterrestrische Physik (MPE), National Astronomical Observatory of China, New Mexico State University, New York University, University of Notre Dame, Observatario Nacional/MCTI, The Ohio State University, Pennsylvania State University, Shanghai Astronomical Observatory, United Kingdom Participation Group, Universidad Nacional Autonoma de Mexico, University of Arizona, University of Colorado Boulder, University of Oxford, University of Portsmouth, University of Utah, University of Virginia, University of Washington, University of Wisconsin, Vanderbilt University, and Yale University.

This project makes use of the MaNGA-Pipe3D dataproducts. We thank the IA-UNAM MaNGA team for creating this catalogue, and the Conacyt Project CB-285080 for supporting them.

%


\software{astropy \citep{collaboration2013astropy,astropycollaboration2018astropy}, matplotlib  \citep{hunter2007matplotlib}
          }




\bibliography{mzrhi.bib}{}
\bibliographystyle{aasjournal}



\end{document}